\documentclass[conference]{IEEEtran}
\IEEEoverridecommandlockouts
\usepackage{cite}
\usepackage{amsmath,amssymb,amsfonts}
\usepackage{algorithmic}
\usepackage{graphicx}
\usepackage{textcomp}
\usepackage{xcolor}
\usepackage{amsmath,lipsum}
\usepackage[ruled]{algorithm2e}
\usepackage{graphicx}
\usepackage{float}
\usepackage{subfigure}
\usepackage{multirow}
\usepackage{multicol}
\usepackage{arydshln}
\usepackage{cite}
\usepackage{stfloats}

\def\BibTeX{{\rm B\kern-.05em{\sc i\kern-.025em b}\kern-.08em
    T\kern-.1667em\lower.7ex\hbox{E}\kern-.125emX}}
\begin{document}

\title{Distributed Deep Learning Inference Acceleration using Seamless Collaboration in Edge Computing\\ }
\author{\IEEEauthorblockN{Nan~Li, Alexandros~Iosifidis and Qi~Zhang }
\IEEEauthorblockA{DIGIT, Department of Electrical and Computer Engineering, Aarhus University.\\
Email: \{linan, ai, qz\}@ece.au.dk}}
\maketitle

\begin{abstract}
This paper studies inference acceleration using
distributed convolutional neural networks (CNNs) in collaborative edge computing. To ensure inference accuracy in inference task partitioning, we consider the receptive-field when performing segment-based partitioning. To maximize the parallelization between the communication and computing processes, thereby minimizing the total inference time of an inference task, we design a novel task collaboration scheme in which the overlapping zone of the sub-tasks on secondary edge servers (ESs) is executed on the host ES, named as HALP. We further extend HALP to the scenario of multiple tasks. Experimental results show that HALP can accelerate CNN inference in VGG-16 by 1.7-2.0x for a single task and 1.7-1.8x for 4 tasks per batch on GTX 1080TI and JETSON AGX Xavier, which outperforms the state-of-the-art work MoDNN. Moreover, we evaluate the service reliability under time-variant channel, which shows that HALP is an effective solution to ensure high service reliability with strict service deadline.
\end{abstract}

\begin{IEEEkeywords}
Distributed CNNs, Receptive-field, Edge computing, Inference acceleration, Service reliability, Delay constraint
\end{IEEEkeywords}

\section{Introduction}
Recently deeper convolutional neural networks (CNNs) based intelligent IoT applications are becoming more and more popular in the emerging applications, such as video surveillance and industrial intelligent control. However, IoT devices are often not able to run computation-intensive tasks locally and meet stringent latency requirements \cite{Mao2017Modnn}. To address this issue, various light-weight CNN architectures are designed to reduce multiply-accumulate operations; however, these methods inevitably cause loss of inference accuracy \cite{Xu2020}. 

To meet the service latency requirements of time-critical IoT applications, a promising approach is edge computing, which enables IoT devices to offload computation intensive tasks to edge server (ES) in their proximity \cite{Jianhui2020IoT,Qi2015IoT}. However, the fluctuations in offloading time caused by the stochastic offloading channel state, may result in missing service deadline \cite{Nan2017CSI,jianhui2020Access}. Accelerating CNN inference is a feasible approach to address the
uncertainties in offloading time, thereby ensuring high service reliability for time-critical IoT applications. 

In our proposed collaborative edge computing for distributed CNN inference, the IoT device offloads an inference task to the host ES, which further partitions the task and distributes each sub-task to individual secondary ES. The inference task is partitioned based on the fundamental processing characteristics of CNN. As we know, for a convolutional layer (CL), the input is a tensor with the shape (height, width, channel). For instance, the input of the first CL is the original image with 3 channels (R, G, B). The convolution operation of a CL is basically a {\textit{dot product}} between the kernel-sized patch of the input and the kernel, which is then summed to generate one entry of the output tensor. Each kernel will apply this operation across the spatial dimensions (width and height) of the input tensor to generate a 2-dimensional feature map of the output tensor. Theoretically speaking, the computation of a CL can be partitioned and processed by different computation units collaboratively without loss of inference accuracy. However, how to efficiently partition, distribute and schedule an inference task within a cluster of ESs is a challenging problem that has not been adequately addressed.

To partition an inference task, segment-based spatial partitioning are proposed to partition the output tensor along the largest dimension into 1-dimensional segmentation \cite{Hsu2020ICC,Mohammed2020Infocom}. However, the above methods ignored the effect of stride and padding for each convolutional layer (CL), which may potentially cause loss of accuracy. To address this issue, we consider the receptive-field \cite{luoreceptive} when using segment-based partitioning to partition an inference task, which will not compromise any inference accuracy.

To allow multiple ESs to process sub-tasks simultaneously, layer-wise parallelization was used in the previous works \cite{Mao2017Modnn,Mohammed2020Infocom}. However, in these works, the host ES needs to receive and merge the sub-outputs from all the secondary ESs for every CL, then again partitions the output tensor into multiple sub-tasks and transmits them to each secondary ES as the new sub-input. This will result in substantial communication time \cite{Jianhui2020IoT}. To minimize the communication time, we propose a novel host ES assisted layer-wise parallelization (HALP) method, in which the overlapped part of two sub-tasks (referred to as {\textit{overlapping zone}} in this paper) is executed at the host ES. In this way, each secondary ES only needs to receive the needed information to process the next CL from the host ES which is very small, for example, only 4 rows are need in order to proceed with the 2nd CL of VGG-16. The salient benefit of this is that the communication time is so small that it can be done, while the ESs are in the process of computing. In other words, such communication processes do not take up time (or take up very little) of the overall inference time. 

This paper studies inference acceleration of distributed CNNs leveraging seamless collaboration in edge computing to minimize the gaps between communication and computation. Our contributions are summarized as follows. 
\begin{itemize}
\item We design a novel task collaboration scheme, HALP, to maximize the parallelization between communication and computing processes, thereby minimizing the total inference time of an inference task. We further extend HALP to the scenario of multiple inference tasks.
\item To measure the performance of HALP, we conduct experiments on the high-end GPU GTX 1080TI and the embedded GPU JETSON AGX Xavier. The experimental results show that HALP can accelerate CNN inference of VGG-16 by 1.75-2.04x for a single task and 1.67-1.81x for 4 tasks per batch, which outperforms the state-of-the-art layer-wise parallelization MoDNN \cite{Mao2017Modnn}.
\item We evaluate the service reliability under time-variant channel, which shows that HALP is effective to ensure high service reliability with strict service deadline.
\end{itemize}

The remainder of this paper is organized as follows. Section \ref{background} provides a background of CNNs and how to calculate the receptive-field. Section \ref{systemmodel} presents the system model of our proposed distributed CNN inference using collaborative ESs. In Section \ref{HALPCNN}, we design the novel task collaboration scheme HALP. The simulation results are presented and discussed in Section \ref{simulation}, and the conclusions are drawn in Section \ref{conclusion}. The source code will be made publicly available at https://gitlab.au.dk/netx/agileiot/halp.git.
\section{Background}\label{background}
In CNNs, every kernel looks at a specific part of the input tensor (named as \textit{Receptive-Field} \cite{luoreceptive}), performs multiplication-addition operations to generate an entry of the output tensor, and then moves by a defined number of pixels ({\textit{stride}}), as shown in Fig. \ref{Fig.receptive}.
In this study, the input image of CNNs has equal height and width. We denote the height of the input and output tensor of CL $g_i$ as $I_{g_i}$ and $O_{g_i}$ respectively. Assuming the kernel size, the padding size and the stride size of CL $g_i$ are $k_{g_i}$, $p_{g_i}$ and $s_{g_i}$ respectively, the attributes of the output tensor can be calculated as \cite{034}
\begin{equation}
	{\textit{O}_{g_i}} = \left\lfloor  {\left( I_{g_i} + 2p_{g_i} - k_{g_i}\right)/s_{g_i} } \right\rfloor  + 1,
	\label{eq1}
\end{equation}
\begin{equation}
	{j_{g_i}} = {j_{g_{i-1}}}  s_{g_i},
	\label{eq2}
\end{equation}
\begin{equation}
	{r_{g_i}} = {r_{g_{i-1}}} + \left( {k_{g_i} - 1} \right) {j_{g_{i-1}}},
	\label{eq3}
\end{equation}
\begin{equation}
	\sigma_{g_i} = \sigma_{g_{i-1}} + \left[ \left(k_{g_i} - 1 \right)/2 - p_{g_i} \right ] {j_{g_{i-1}}},
	\label{eq4}
\end{equation}
where $j_{g_i}$ is the cumulative stride (referred to as \textit{jump}) of each output pixel, $r_{g_i}$ is the receptive-field size of each output pixel, and $\sigma_{g_i}$ is both the row index and column index of the center position of the receptive-field of the first element in the output tensor. Note that the input tensor of CL $g_i$ is the output tensor of CL $g_{i-1}$, that means, ${\textit{I}_{g_i}} = {\textit{O}_{g_{i-1}}}$.
\section{System Model}\label{systemmodel}
\begin{figure}[h]
\begin{minipage}[t]{0.5\textwidth}
\centering
\includegraphics[width=0.7\textwidth]{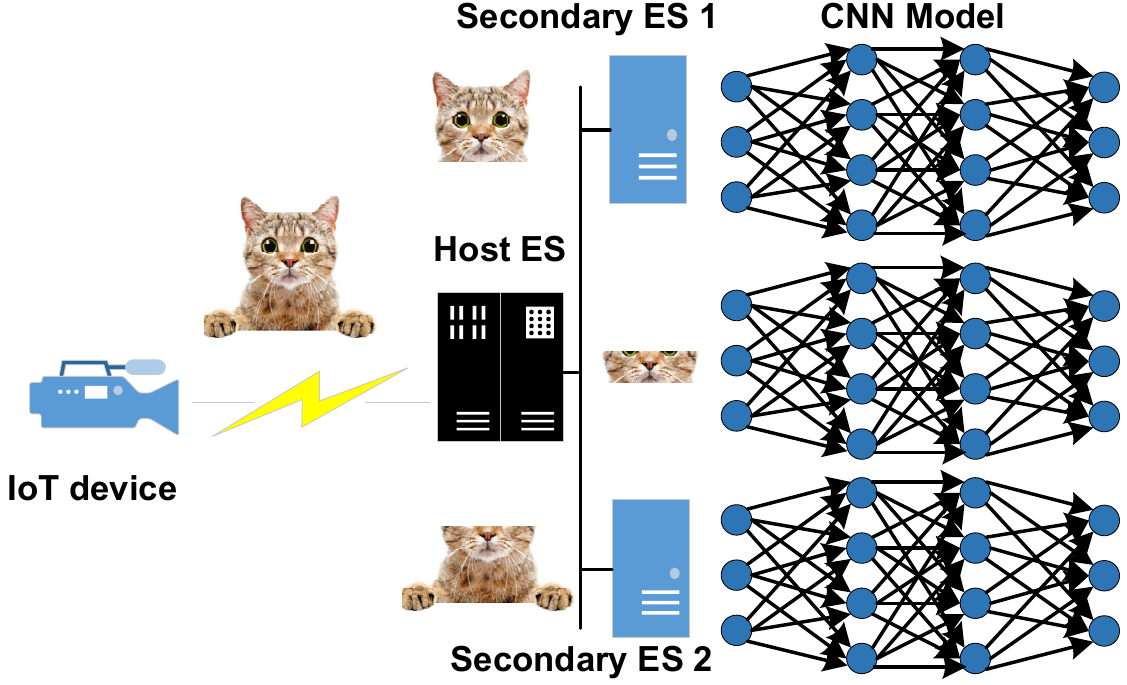}
\vspace{-1.5mm}
\caption{An example of the collaborative edge computing for an inference task}
\label{Fig.systemmodel}
\vspace{1mm}
\end{minipage}
\begin{minipage}[t]{0.5\textwidth}
\centering
\includegraphics[width=0.65\textwidth]{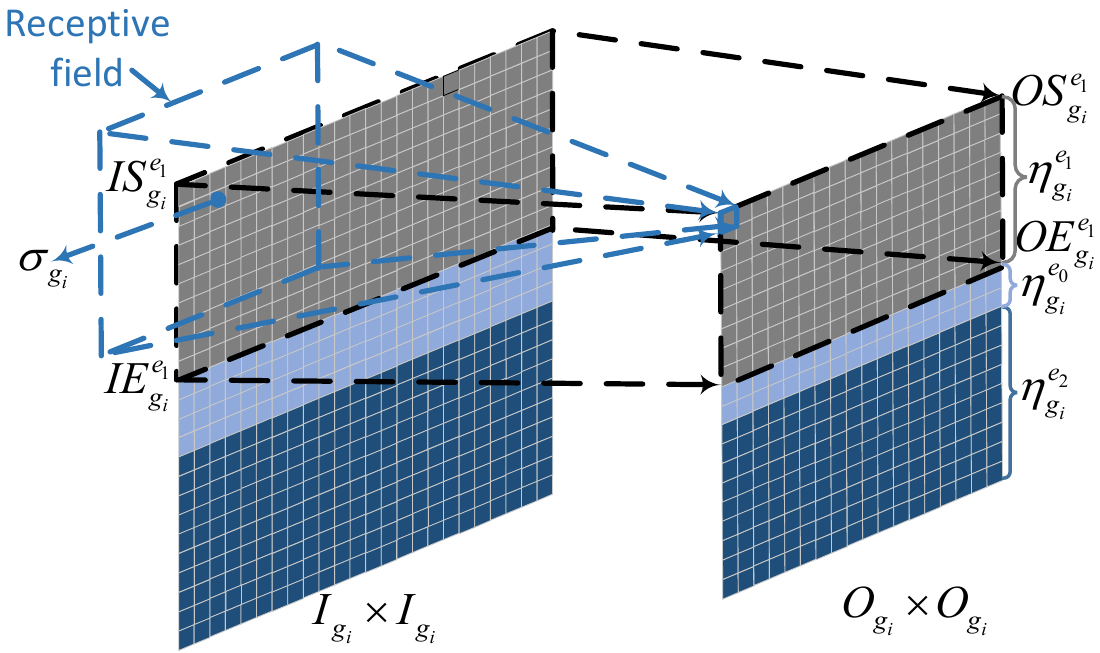}
\vspace{-1.5mm}
\caption{Receptive-field of an output tensor}
\label{Fig.receptive}
\end{minipage}
\vspace{-2mm}
\end{figure}

In this paper we studied inference acceleration using distribute CNNs in collaborative edge computing system. The basic system is composed of one IoT device, a host ES and two secondary ESs. This system can be extended to multiple IoT devices and multiple secondary ESs. In this system, the IoT device can offload CNN inference tasks, e.g., images, to its host ES, which can further partition an inference task and distribute the sub-tasks to multiple secondary ESs and then together complete the task collaboratively. The connection between the host ES and the secondary ESs is high-speed Ethernet. An example of the collaborative inference for a task is shown in Fig. \ref{Fig.systemmodel}, in which the overlapping zone between two sub-tasks is processed on host ES. The inference task is assumed to use a CNN with $N$ CLs and several fully connected layers (FLs), in which the set of CLs is denoted as $ {\bf{G}} = \left\{ {{g_1}, \cdots ,{g_N}}\right\}$. The set of ESs can be denoted as $ {\bf{E}} = \left\{e_0, e_1,e_2 \right\}$, where $e_0$ represents the host ES, $e_1$ and $e_2$ are two secondary ESs. 

The host ES partitions an inference task into sub-tasks using segment-based partitioning \cite{Mohammed2020Infocom}. As shown in Fig. \ref{Fig.receptive}, we define a variable $\eta_{g_i}^{e_k}$ to denote the ratio of the sub-output on ES $e_k$ to the entire output tensor of CL $g_i$ which is subject to
\begin{equation}
		\begin{array}{*{20}{c}}
		\sum\limits_{e_k \in {\bf{E}}} {\eta _{g_i}^{e_k}} = 1,
		\end{array}
		\label{eq1}
\end{equation}
where $0 \leq \eta _{g_i}^{e_k}\leq 1$, and $\eta_{g_i}^{e_0}$ is the ratio of the sub-output of the overlapping zone.

Therefore, the start and end row index of the sub-output on ES $e_k$, $\textit{OS}_{g_i}^{e_k}$ and $\textit{OE}_{g_i}^{e_k}$, can be computed as follows,
\begin{eqnarray}
	{\textit{OS}}_{g_i}^{e_k} &=& \left\{ {\begin{array}{*{20}{c}}
		{1},& e_k = e_1,\\
		{\eta_{g_i}^{e_1} O_{g_i} + 1},&e_k = e_0,\\
		{\left(\eta_{g_i}^{e_1}+\eta_{g_i}^{e_0} \right) O_{g_i} + 1},&e_k = e_2.
	\end{array}} \right. \\
	{\textit{OE}}_{g_i}^{e_k} &=& \left\{ {\begin{array}{*{20}{c}}
		{\eta_{g_i}^{e_1} O_{g_i}},\;\;\;\;\;\;\;&e_k = e_1,\\
		{\left(\eta_{g_i}^{e_1}+\eta_{g_i}^{e_0}\right)} O_{g_i},\;\;\;\;\;\;\;&e_k = e_0,\\
		{O_{g_i}},\;\;\;\;\;\;\;&e_k = e_2. \end{array}} \right.
\end{eqnarray}
To ensure the inference accuracy, we calculate the start row index and end row index of sub-input on ES $e_k$, $\textit{IS}_{g_i}^{e_k}$ and $\textit{IE}_{g_i}^{e_k}$, based on the receptive field of CL $g_i$ in Section \ref{background}, as follows,
\begin{equation}
	{\textit{IS}}_{g_i}^{e_k} \hspace{-1mm}=\hspace{-1mm} \max \left(\sigma_{g_i} +  \left({\textit{OS}}_{g_i}^{e_k}-1\right) j_{g_i} -\left\lfloor {\left(r_{g_i}-1\right)/2} \right\rfloor,1 \right).
		\label{eq4}
\end{equation}
\begin{equation}
	{\textit{IE}}_{g_i}^{e_k} \hspace{-1mm}=\hspace{-1mm} \min \left(\sigma_{g_i} +  \left({\textit{OE}}_{g_i}^{e_k}+1\right) j_{g_i} -\hspace{-1mm}\left\lfloor {\left(r_{g_i}-1\right)/2} \right\rfloor,I_{g_i} \right).
		\label{eq5}
\end{equation}
\section{Host Assisted layer-wise parallelization for distributed CNN inference} \label{HALPCNN}
\begin{table}[]
\centering
\caption{Inference time of a single task on GTX 1080TI at 40 G{\upshape bps} transmission data rate between ESs ({\upshape ms})}
\scalebox{0.85}{
\begin{tabular}{|c|c|c|c|c|c|c|c|}
\hline
\multirow{2}{*}{} & \multicolumn{4}{c|}{Secondary ES $e_k$} & \multicolumn{3}{c|}{Host ES $e_0$} \\ \cline{2-8} & \multicolumn{1}{c|}{} & \multicolumn{1}{c|}{} & \multicolumn{1}{c|}{} & \multicolumn{1}{c|}{} & \multicolumn{1}{c|}{}& \multicolumn{1}{c|}{}& \multicolumn{1}{c|}{}\\[-0.8em]
                  & $t_{\textit{int}}^{e_0\rightarrow e_k}$      & $t_{\textit{cmp}}^{e_k\rightarrow e_0}$      & $t_{\textit{com}}^{e_k\rightarrow e_0}$   & $t_{\textit{cmp}}^{e_k}$  & $t_{\textit{cmp}}^{e_0\rightarrow e_k}$          & $t_{\textit{com}}^{e_0\rightarrow e_k}$    & $t_{\textit{cmp}}^{e_0}$         \\[1pt] \hline
CL $g_1$              & 0.057   & 0.001   & 0.011       &  0.044     & 0.001               &  0.011 &  0           \\ \hline
CL $g_2$               & 0      &0.002          &0.011       & 0.094       &  0             &0 & 0.003           \\ \hline
\end{tabular}}
\label{tab:inference}
\vspace{-4mm}
\end{table}
\addtolength{\topmargin}{0.03in}
In this section, we firstly explain how host assisted layer-wise parallelization (HALP) works for a single inference task. Then we apply HALP to the scenario of multiple inference tasks arriving simultaneously. 
\begin{figure*}[htbp]
\centering
\begin{minipage}[t]{0.39\textwidth}
\centering
\includegraphics[width=0.9\textwidth]{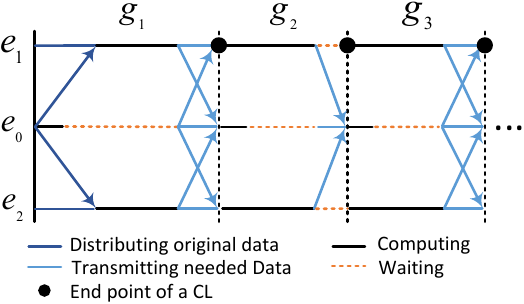}
\vspace{-2mm}
\caption{Conventional layer-wise parallelization for VGG-16}
\label{traditionalschedule}
\end{minipage}
\begin{minipage}[t]{0.56\textwidth}
\centering
\includegraphics[width=0.93\textwidth]{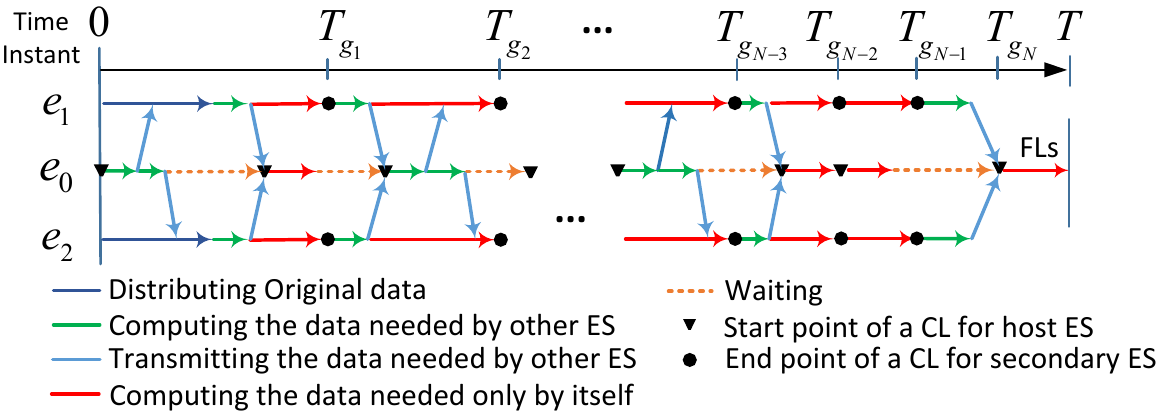}
\centering
\vspace{-2mm}
\caption{Communication and computation processes of HALP for VGG-16}
\label{one_task}
\end{minipage}
\vspace{-5mm}
\end{figure*}
\subsection{Host Assisted layer-wise parallelization}\label{HALP}
The conventional layer-wise parallelization using collaborative edge computing for VGG-16 is shown in Fig. \ref{traditionalschedule}. Basically, the host ES and the secondary ESs exchange the needed data after computing each CL, thus, communication and computing are two separate processes in the timeline. Therefore, the inference time of each ES is the sum of communication time, computation time and waiting time. We denote the communication time that distributing the original image data from host ES $e_0$ to secondary ES $e_k$ by $t_{\textit{int}}^{e_0\rightarrow e_k}$, the time to compute data that is later needed at $e_j$ but computed at $e_i$ by $t_{\textit{cmp}}^{e_i \rightarrow e_j}$, the time of sending the data from $e_i$ to $e_j$ by $t_{\textit{com}}^{e_i \rightarrow e_j}$, and the time to compute the data only needed by $e_k$ by $t_{\textit{cmp}}^{e_k}$. We test the inference time of VGG-16 on GTX 1080TI assuming the data rate between ESs is 40 Gbps. The total inference time of VGG-16 is 3.142 ms, the inference time of $g_1$ and $g_2$ are 0.113 ms and 0.107 ms respectively. Table \ref{tab:inference} shows the detailed inference time of $g_1$ and $g_2$. We can see the time of transmitting the data between ESs for $g_1$ and $g_2$ take up about 60.2\%, i.e., $(0.057+0.011)/0.113$, and 10.3\%, i.e., $0.011/0.107$, of their inference time. 

The key observation is that the communication and computation for each CL can in fact be processed in parallel. This is because the time to transmit the data needed by another ES is much smaller compared with the computing time. Thus, the communication can in a large degree get done, while the computing is on going. We design HALP to parallelize the process of data communication with the process of computation as much as possible for each CL, as shown in Fig. \ref{one_task}. In this way, we can minimize the total inference time. The basic idea of HALP is as follows. Before starting the computation of the first CL $g_1$, the host ES $e_0$ will first partition an image into three parts. Among them, two parts are of size close to half of the image and each will be sent to a collaborating secondary ES. The size of data distributed to secondary ES $e_k \in \left\{e_1,e_2\right\}$ can be calculated as
\begin{equation}
S_{\textit{int}}^{e_0\rightarrow e_k} = 
				4{\left( {{\textit{IE}}_{g_1}^{e_k} - {\textit{IS}}_{g_1}^{e_k} + 1} \right) {\textit{I}}_{g_1} c_{g_1}^{\textit{in}}},	e_k \in \left\{e_1,e_2\right\}. 
\label{eq6}
\end{equation}
where $c_{g_i}^{\textit{in}}$ is the number of channels of the input tensor of $g_i$. Note that for a tensor $\left(\textit{height}, \textit{width},\textit{channel} \right)$ with type in float32, its size is calculated as $4\cdot \textit{height} \cdot \textit{width} \cdot \textit{channel}$ in bytes. The third part (overlapping zone) will be processed by host ES itself, which is commonly very small (depending on the attributes of a CL). For example, the overlapping zone is only 4 rows of pixels for VGG-16 \cite{Karen2015VGG}, because the kernel size used in VGG-16 is 3.

While the host ES $e_0$ transmits parts of the original input image to the secondary ESs, it computes the output of the overlapping zone of the first CL. Then host ES sends the corresponding output of $g_1$ to the secondary ESs as soon as it is available. For each secondary ES $e_k$, after receiving the part of the original image, it first computes the part of output of $g_1$ needed by the host ES to proceed with computing CL $g_2$. The secondary ES transmits this part of the output to the host ES, while it is computing the rest output of $g_1$ needed only by itself. Such kind of seamless collaboration between host ES and secondary ESs will repeat for every CL from $g_1$ until $g_{N}$. Note that in VGG-16, if the next layer is pooling layer, the host ES does not need to send the output of the current CL to secondary ESs because the secondary ESs already have the information to proceed with the pooling layer. For example, in Fig. \ref{one_task}, after completing the computation of $g_2$, only the host ES $e_0$ gets the needed data from secondary ES $e_k$ to proceed with $g_3$. For the last CL $g_N$, all the sub-outputs of the secondary ESs will be sent to host ES and then be merged as the input for FLs. 

Therefore, the size of data sent from host ES $e_0$ to secondary ES $e_k \in \left\{e_1,e_2\right\}$ and from secondary ES $e_k$ to $e_0$ of $g_i$, $S_{g_i}^{e_0\rightarrow e_k}$ and $S_{g_i}^{e_k\rightarrow e_0}$, can be expressed as follows,
\begin{equation}
		S_{g_i}^{e_0\rightarrow e_1} = 
				4{\left( {{\textit{IE}}_{g_i}^{e_1} - {\textit{OS}}_{g_{i-1}}^{e_0} + 1} \right) {\textit{I}}_{g_i} c_{g_i}^{\textit{in}}},  \;\;\;g_i \ne g_N.
\end{equation}
\begin{equation}
		S_{g_i}^{e_0\rightarrow e_2} = 
				4{\left( {{\textit{IS}}_{g_i}^{e_2} - {\textit{OE}}_{g_{i-1}}^{e_0} + 1} \right) {\textit{I}}_{g_i} c_{g_i}^{\textit{in}}},\;\;\;g_i \ne g_N.
\end{equation}
\begin{equation}
		S_{g_i}^{{e_1}\rightarrow e_0} = \left\{ {\begin{array}{*{20}{c}}
		\hspace{-2mm}{4{\left( {\textit{IS}}_{g_{i+1}}^{e_0}-{\textit{OE}}_{g_i}^{e_1} + 1 \right) {\textit{I}}_{g_i} c_{g_i}^{\textit{in}}} },&  g_i \ne g_N ,\\
		\hspace{-2mm}{4{\left( {{\textit{OE}}_{g_i}^{e_1} - {\textit{OS}}_{g_i}^{e_1} + 1} \right) {\textit{O}}_{g_i} c_{g_i}^{\textit{out}}}}, & g_i = g_N.
		\end{array}} \right.
\end{equation}
\begin{equation}
		S_{g_i}^{{e_2}\rightarrow e_0} =  \left\{ {\begin{array}{*{20}{c}}
		\hspace{-2mm}{4{\left( {{\textit{IS}}_{g_i}^{e_2} - {\textit{OE}}_{g_{i+1}}^{e_0} + 1} \right) {\textit{I}}_{g_i} c_{g_i}^{\textit{in}}} },&  g_i \ne g_N ,\\
		\hspace{-2mm}{4{\left( {{\textit{OE}}_{g_i}^{e_2} - {\textit{OS}}_{g_i}^{e_2} + 1} \right) {\textit{O}}_{g_i} c_{g_i}^{\textit{out}}}}, & g_i = g_N.
		\end{array}} \right.
\end{equation}
\setcounter{equation}{18}
\begin{figure*}[hb]
\begin{equation}		
    T_{g_i}^{e_0} = \left\{ {\begin{array}{*{20}{c}}
			{\max \left( {t_{g_i}^{e_0},\max\limits_{e_k \in {\bf{E}}\setminus{e_0}} {\left(t_{{\textit{int}}}^{e_0 \rightarrow e_k} + t_{g_i,{\textit{cmp}}}^{e_k \rightarrow e_0} +t_{g_i,{\textit{com}}}^{e_k \rightarrow e_0}\right)}}\right)},&{g_i = g_1},\\
			{\max \left( t_{g_i}^{e_0} + T_{g_{i-1}}^{e_0},\max\limits_{e_k \in {\bf{E}}\setminus{e_0}}{\left(T_{g_{i-1}}^{e_k} + t_{g_i,{\textit{cmp}}}^{e_k \to {e_0}} + t_{g_i,{\textit{com}}}^{e_k \to {e_0}}\right)} \right)},&g_i \in {\bf{G}}\setminus \left\{g_i,g_N\right\},\\
			{\max \left( t_{g_i}^{e_0} + T_{g_{i-1}}^{e_0},\max\limits_{e_k \in {\bf{E}}\setminus{e_0}}{\left(T_{g_{i-1}}^{e_k} + t_{g_i,{\textit{cmp}}}^{e_k} + t_{g_i,{\textit{com}}}^{e_k \to {e_0}}\right)} \right)},&{g_i = g_N}.
	\end{array}} \right.
\end{equation}
\vspace{-1mm}
\end{figure*}
\addtolength{\topmargin}{0.08in}
Assuming the start time instant of CL $g_1$ is 0 and the end time instant of CL $g_i$ is $T_{g_i}$, the objective to minimize the inference time can be denoted as 
\setcounter{equation}{14}
\begin{equation}
		\mathop {\min } {T_{g_N}+ t_{{\textit{FLs}}}},
		\label{eq16}
\end{equation}
where $t_{\textit{FLs}}$ is the inference time of FLs.

The inference time that the secondary ES $e_k \in \left\{e_1,e_2\right\}$ spends on CL $g_i$ can be denoted as
\begin{equation}
	t_{g_i}^{e_k} = \left\{ {\begin{array}{*{20}{c}}
	{t_{{\textit{int}}}^{e_0 \rightarrow e_k} + 				t_{g_i,{\textit{cmp}}}^{e_k \rightarrow e_0} +\max \left(	t_{g_i,{\textit{com}}}^{e_k \rightarrow e_0}, t_{g_i,{\textit{cmp}}}^{e_k} \right)},&g_i = g_1, \\
			{t_{g_i,{\textit{cmp}}}^{e_k \rightarrow e_0} +\max \left(t_{g_i,{\textit{com}}}^{e_k \rightarrow e_0},t_{g_i,{\textit{cmp}}}^{e_k } \right)},&{g_i \ne g_1}.
			\end{array}} \right.
	\label{eq18}
\end{equation}
The end time instant of secondary ESs $e_k \in \left\{e_1,e_2\right\}$ for CL $g_i$ can be denoted as
\begin{equation}
		T_{g_i}^{e_k} = \left\{ {\begin{array}{*{20}{c}}
				t_{g_i}^{e_k}, &{g_i = g_1},\\
				t_{g_i}^{e_k} + T_{g_{i-1}}^{e_k},&{\rm{others}}. 
		\end{array}} \right.
		\label{eq19}
\end{equation}
Correspondingly, the inference time that host ES $e_0$ spends on CL $g_i$ can be denoted as
\setcounter{equation}{17}
\begin{equation}
			t_{g_i}^{e_0} = \left\{ {\begin{array}{*{20}{c}}
					{ t_{g_i,{\textit{cmp}}}^{e_0 \rightarrow e_1} +\max \left(t_{g_i,{\textit{com}}}^{e_0 \rightarrow e_1}, t_{g_i,{\textit{cmp}}}^{e_0 \rightarrow e_2} +t_{g_i,{\textit{com}}}^{e_0 \rightarrow e_2}\right)}, & g_i \ne g_N ,\\
					t_{g_i,{\textit{cmp}}}^{e_0},&g_i = g_N.
			\end{array}} \right.
	\label{eq20}
\end{equation}
The end time instant of host ES $e_0$ for CL $g_i$ is denoted as (19), below. Therefore, the end time instant of CL $g_i$, $T_{g_i}$, can be denoted as,
	\setcounter{equation}{19}
	\begin{equation}
		T_{g_i} = \max \left( T_{g_i}^{e_0},T_{g_i}^{e_1},T_{g_i}^{e_2} \right).
	\end{equation}
	
To evaluate the acceleration performance of the proposed
HALP compared with the approach of the pre-trained model running on a standalone ES, we define the speedup ratio $\rho $ as
	\begin{equation}
		\rho= 1-\left(T_{g_N}+ t_{\textit{FLs}}\right)/t_{\textit{pre}}.
	\end{equation}
where $t_{\textit{pre}}$ is the CNN inference time for an inference task running on a standalone ES.
\subsection{HALP for multiple inference tasks}
\begin{figure}[h]
\vspace{-5mm}
		\centering
	\includegraphics[width=0.28\textwidth]{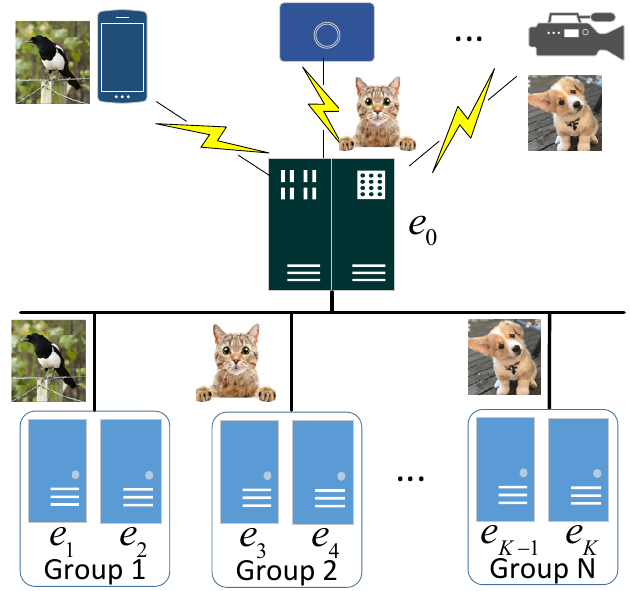}		
	\caption{HALP for multiple tasks}
	\label{Fig.multiple_task}
	\vspace{-3mm}
	\end{figure}

In the HALP scheme, a host ES has a much smaller sub-task compared with a secondary ES. The rationale behind this design is that in a practical system there are more than one IoT device. We assume a host ES can provide offloading service to multiple IoT devices and then distribute these tasks to a cluster of secondary ESs and manage the collaborative edge computing of the tasks. In other words, the host ES can process the sub-tasks for multiple tasks and every two secondary ESs form a collaborating group, as shown in Fig. \ref{Fig.multiple_task}. We assume the host ES processes the overlapping zones for multiple tasks sequentially, e.g., the overlapping zone for task 1 is computed first.

Assume the set of ESs for multiple tasks is $E^{*}=\{e_0,e_1,e_2,\cdots,e_K \}$, the inference time that host ES $e_0$ spends on CL $g_i$ in (\ref{eq20}) can be updated as 
\begin{equation}
		t_{g_i}^{e_0} = \left\{ {\begin{array}{*{20}{c}}
				\hspace{-2mm}{ \max\limits_{e_k \in {\bf{E^{*}}}\setminus{e_0}} \left( \sum\limits_{j=1}^{k} {t_{g_i,{\textit{cmp}}}^{e_0 \rightarrow e_j}} + t_{g_i,{\textit{com}}}^{e_0 \to e_k}\right)}, & g_i \ne g_N,\\
		t_{g_i,{\textit{cmp}}}^{e_0},&g_i = g_N.
		\end{array}} \right.
		\label{eq24}
\end{equation}
Correspondingly, for multiple tasks, the end time instant of host ES $e_0$ for CL $g_i$ in (19) will be updated by using $\textbf{E}^{*}$ to replace $\textbf{E}$. Therefore, the end time instant of CL $g_i$ for multiple tasks, $T_{g_i}$ in (20), can be updated as,
\begin{equation}
	T_{g_i}= \max \left( T_{g_i}^{e_0},T_{g_i}^{e_1},\cdots,T_{g_i}^{e_K}\right).
	\label{eq26}
\end{equation}
\begin{figure*}[htbp]
\centering
\begin{minipage}[t]{0.48\textwidth}
\centering
\includegraphics[width=0.6\textwidth]{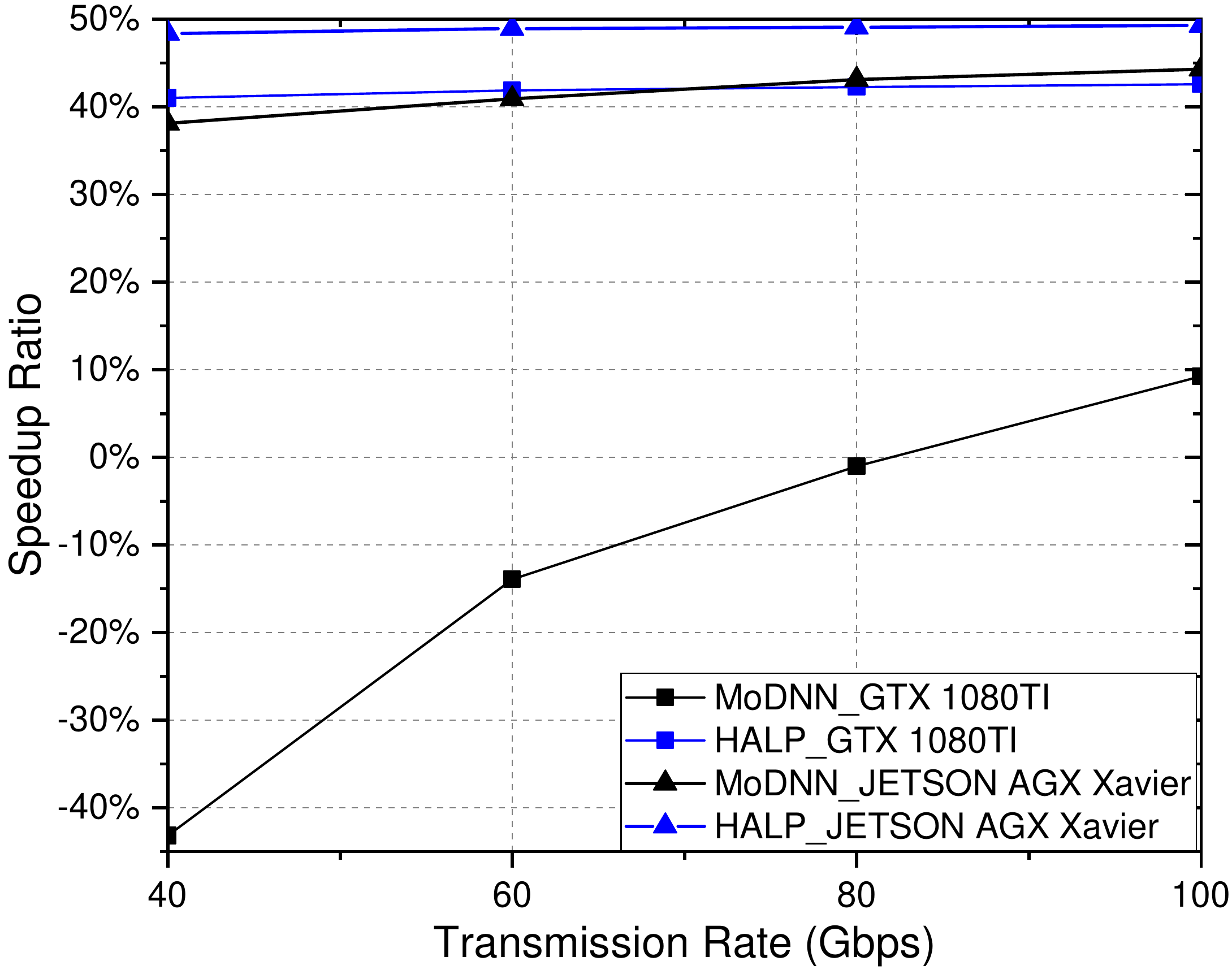}
\vspace{-2mm}
\caption{Speedup ratio of the scenario with single inference task}
\label{Fig.trans_one_task}
\end{minipage}
\begin{minipage}[t]{0.48\textwidth}
\centering
\includegraphics[width=0.6\textwidth]{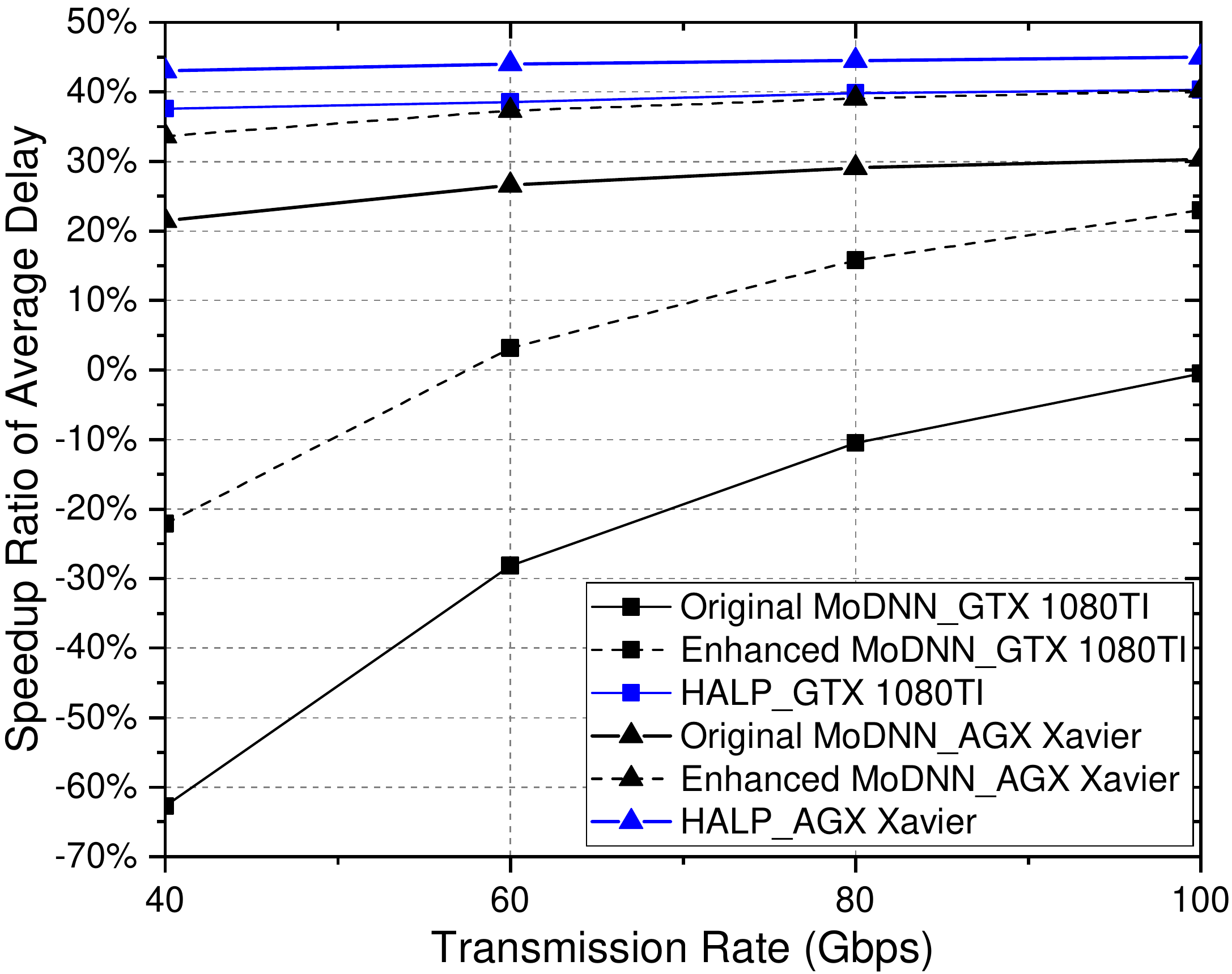}
\vspace{-2mm}
\caption{Speedup ratio of 4 inference tasks under average delay}
\label{Fig.trans_multiple_task}
\end{minipage}
\vspace{-5mm}
\end{figure*}
\section{Performance Evaluation}\label{simulation}
\subsection{Simulation Setup and Methodology}
We conduct simulations in a collaborative edge computing network consisting of 9 ESs with equal computing capacity. We feed 4 inference tasks into the system at a time. The data transmission between ESs is via Ethernet with transmission rate ranging ranging from 40 Gbps to 100 Gbps \cite{IEEE2021}. We use the image classification model VGG-16 \cite{Karen2015VGG} in the simulation. To evaluate the performance
of the proposed HALP, we compare it to a state-of-the-art layer-wise parallelization MoDNN \cite{Mao2017Modnn}. To provide a realistic evaluation of the performance using different types of ESs, we run the distributed CNN on GTX 1080TI with 11.3 TFLOPS of float32, and JETSON AGX Xavier with 1.3 TFLOPS of float32 in our simulation. The input size of images is $224 \times 224 \times 3$, and the performance metric is speedup ratio $\rho $ and throughput (in frame per second).
\addtolength{\topmargin}{0.06in}
\subsection{CNN inference acceleration of single inference task}
Fig. \ref{Fig.trans_one_task} shows the speedup ratio of one inference task at variable transmission rate between ESs.
It can be seen that HALP can speedup the CNN inference by 2.0x on JETSON AGX Xavier and 1.7x on GTX 1080TI, which is better than MoDNN using the same platform at the given transmission rate. This is mainly due to the fact that HALP can to a large degree parallelize the process of data communication with the process of computation; however, in MoDNN all secondary ESs need to send back the sub-outputs to host ES, then receive the sub-input from host ES for the next CL. Moreover, for the platform of higher computation capacity, i.e., GTX 1080 TI, the speedup ratio difference between HALP and MoDNN is larger. The reason is that a platform with higher computation capacity has smaller computing time, thus, for a given communication time, it takes up more in the overall inference time. In other words, the saved data communication time has a big impact on the overall inference time. Similarly, the difference of speedup ratio between HALP and MoDNN is larger at a lower transmission rate. This is because HALP can save more communication time when transmission rate between ESs is low. 
\subsection{CNN inference acceleration of multiple inference tasks}\label{4task}
In this subsection, we study how HALP performs for a scenario that multiple inference tasks arrive simultaneously. We assume a collaborative Edge Computing system consists of a host ES which coordinates 4 pairs of secondary ESs (i.e., 8 ESs) and 4 inference tasks arrive at a time. Assume the inference time of one task using HALP  (using one pair of ESs) is $T_H$, the total inference time of 4 tasks is close to $T_H$. Therefore, the average delay of each task in HALP will be $T_H$. This is because multiple tasks are computed in parallel by independent pairs of ESs plus coordinated by the host ES in HALP and will be completed nearly at the same time. In the original design of MoDNN (referred to as {\textit{Original MoDNN}}), all computing resources are used to accelerate the inference of one task (i.e., one task is processed by the 9 ESs), thus, if the inference time of one task is $T_M$, the average delay of each task is $2.5T_M$. However, the Original MoDNN cannot utilize the resource in the best way. We made an improvement in MoDNN to better utilize the computation resource (referred to as {\textit{Enhanced MoDNN}}). We first let 3 tasks be processed by MoDNN in parallel using the 9 ESs (i.e., 3 ESs working on one task) and the remaining task is processed by the 9 ESs. Assume the inference time of the first 3 tasks and the last task of Enhanced MoDNN are $T_M^{E_1}$ and $T_M^{E_2}$, the average delay of each task is $T_M^{E_1}+0.25T_M^{E_2}$. We take an example to calculate the speedup ratio under average delay and the average throughput. In our experiment, the inference time $T_H$, $T_M$, $T_M^{E_1}$, $T_M^{E_2}$ and $t^{\textit{pre}}$ of GTX 1080TI at the transmission rate between ESs of 100 Gbps are 2.81 ms, 1.89 ms, 3.13 ms, 1.89 ms and 4.7 ms respectively. The speedup ratio of  Enhanced MoDNN at average delay is $\rho = 1- (3.13+1.89*0.25)/4.7 = 0.233$. The average throughput of Enhanced MoDNN is $\left[1/(T_M^{E_1}+T_M^{E_2})\right]*4= 797$ fps. The rest can be done in the same way.
\begin{table}[t]
	\centering
	\caption{Average throughput of 4 inference tasks in one batch ({\upshape fps})}
		\vspace{-2mm}
	\scalebox{0.8}{
	\begin{tabular}{|l|c|c|c|c|}
		\hline
		\multicolumn{1}{|l|}{Transmission rate between ESs}  & 40 Gbps & 60 Gbps & 80 Gbps & \multicolumn{1}{l|}{100 Gbps} \\ \hline
		Pre-trained model\_GTX 1080TI     &\multicolumn{4}{c|}{851}                   \\ \hline
		Original MoDNN\_GTX 1080TI         & 327    & 415    & 479    & 529                         \\ \hline
		 Enhanced MoDNN\_GTX 1080TI        & 498    & 629    & 724    & 797                         \\ \hline
		{\textbf{HALP\_GTX 1080TI}}       & {\textbf{1364}}    & {\textbf{1384}}   & {\textbf{1413}} & {\textbf{1423}}                         \\ \hline \hline
		Pre-trained model\_AGX Xavier   &\multicolumn{4}{c|}{124}                   \\ \hline
		Original MoDNN\_AGX Xavier  & 98    & 105    & 109    & 112                        \\ \hline
		 Enhanced MoDNN\_AGX Xavier  & 138    & 146    & 151    & 152                        \\ \hline
		{\textbf{HALP\_AGX Xavier}}   & {\textbf{219}}    & {\textbf{221}}  & {\textbf{223}}   & {\textbf{225}}                          \\ \hline
	\end{tabular}
	}
	\label{tab1}
\vspace{-5mm}
\end{table}
\begin{table*}[h]
\caption{Service reliability on JETSON AGX Xavier under different time-variant channels (over 99.999\% in bold)}
\label{tab:7}
\centering
\vspace{-2mm}
\scalebox{0.9}{
\begin{tabular}{|c|c|c|c|c|c|c|c|}
\hline
\multirow{2}{*}{} & \multicolumn{2}{c|}{40Mbps} & \multicolumn{3}{c|}{60Mbps}  & \multicolumn{2}{c|}{100Mbps}                                       \\ \cline{2-8} 
                  & \begin{tabular}[c]{@{}c@{}}$\sigma=1$ ms\\ ($\phi=1.2$ Mbps)\end{tabular} & \begin{tabular}[c]{@{}c@{}}$\sigma=5$ ms\\ ($\phi=5.3$ Mbps)\end{tabular} & \begin{tabular}[c]{@{}c@{}}$\sigma=5$ ms\\ ($\phi=11.0$ Mbps)\end{tabular} & \begin{tabular}[c]{@{}c@{}}$\sigma=9$ ms\\ ($\phi=17.3$ Mbps)\end{tabular} & \begin{tabular}[c]{@{}c@{}}$\sigma=14$ ms\\ ($\phi=23.2$ Mbps)\end{tabular} & \begin{tabular}[c]{@{}c@{}}$\sigma=14$ ms\\ ($\phi=51.3$ Mbps)\end{tabular} & \begin{tabular}[c]{@{}c@{}}$\sigma=18$ ms\\ ($\phi=57.4$ Mbps)\end{tabular} \\ \hline
Pre-trained model & 0.815931                                         & 0.571420                                        & \textbf{1}                                       & 0.999934                                         & 0.992992                                          & \textbf{1}                                        & 0.999640                                          \\ \hline
\textbf{HALP}     & \textbf{1}                                      & 0.999104                                         & \textbf{1}                                       & \textbf{1}                                       & 0.999774                                          & \textbf{1}                                        & \textbf{0.999993}                                 \\ \hline
\end{tabular}}
\vspace{-5mm}
\end{table*}

Fig. \ref{Fig.trans_multiple_task} shows the speedup ratio of the scenario of 4 tasks under the average delay. It can be seen that the speedup ratio of 4 tasks at average delay is slightly lower than that of the scenario of a single inference task in Fig. \ref{Fig.trans_one_task}. The reason is that the host ES needs to process multiple overlapping zones, in this way, the secondary ESs need to wait to receive the needed information from host ES at the last several CLs, because the sub-input of secondary ESs becomes smaller along the computation flow of VGG-16. For MoDNN, multiple tasks processed in parallel can reduce the average delay of each task, therefore, Enhanced MoDNN can achieve better performance than Original MoDNN. However, due to large communication time, Enhanced MoDNN cannot outperform HALP. Table \ref{tab1} shows the average throughput of processing 4 inference tasks at a time. It can be seen that the average throughput of HALP is about 2x of Original MoDNN, 1.5x of Enhanced MoDNN and 1.8x of the pre-trained model running on a standalone JETSON AGX Xavier. For the case of GTX 1080TI, HALP's average throughput is approximately 2.7x-4.2x of Original MoDNN, 1.8x-2.7x of Enhanced MoDNN and 1.7x of the pre-trained model running on a standalone ES. The gain of HALP comes from the reduced computing time due to task partitioning and the saved communication time due to parallelization of data communication and computing.
\subsection{Evaluation of service reliability}
In this full offloading scheme, the host ES is responsible for making decisions on partitioning the tasks and distributing the sub-tasks to secondary ESs, which can reduce the computation load on the resource-constrained IoT Devices. Assuming the IoT device offloads multiple tasks to host ES simultaneously, the total tasks completion time can be denoted as $T = T^{\textit{off}} + T^{\textit{inf}}$, where $T^{\textit{off}}$ and $T^{\textit{inf}}$ are offloading time and inference time of multiple inference tasks.

To ensure real-time inference, we can select the minimal transmission rate between IoT device and the host ES according to TABLE \ref{tab1}. For example, to ensure system throughput (e.g., 30 FPS) of the scenario with 4 inference tasks (e.g., each input image of 125 KBytes) processed in parallel, the transmission rate between IoT device and the host ES should not be lower than 31 Mbps. Generally, the channel state is stochastic, which may cause fluctuations in offloading time $T^{\textit{off}}$. In mission critical IoT services, the required service reliability (the probability of meeting the deadline) can be in the range of 99\% to 99.999\%. We assume the offloading time $T^{\textit{off}}\sim\mathcal{N}(\mu,\,\sigma^{2})$. The fluctuation of transmission rate is denoted as $\phi$, which can be estimated based on 3-sigma rule of thumb. TABLE \ref{tab:7} shows the service reliability under different transmission rate. We can see the service reliability will decrease as the fluctuation of channel data rate increases. To address this issue, a higher transmission rate can contribute to improving the service reliability. However, relying solely on improving transmission rate cannot guarantee the service reliability of 99.999\% using the pre-trained model running on a standalone ES. For example, the pre-trained model running on a standalone ES can achieve max 99.964\% service reliability at average data rate of 100 Mbps and fluctuation $\phi = 57.4$ Mbps. In this case, HALP can still ensure high service reliability of 99.999\% and even further. This is because HALP can significantly reduce the inference time as shown in Section \ref{4task}, by which it gives the system more time budget to transmit the image and meet the deadline.
\section{Conclusion}\label{conclusion}
This paper studies inference acceleration of distributed CNNs by leveraging seamless collaboration in edge computing. To ensure the inference accuracy in task partitioning, we use the receptive-field in segment-based partitioning. To minimize the total inference time, we design a novel task collaboration scheme, HALP, to maximize the parallelization between communication and computing processes. We further apply HALP to the scenario of multiple tasks. Experimental results show that HALP can accelerate CNN inference of VGG-16 by 1.7-2.0x for a single task and 1.7-1.8x for 4 tasks per batch on GTX 1080TI and JETSON AGX Xavier, which outperforms the state-of-the-art work MoDNN. Moreover, we evaluate the service reliability under time-variant channel, which shows that HALP is an effective solution to ensure high service reliability with strict service deadline. In our future work, we will study the effect of ESs with diverse performance.
\ifCLASSOPTIONcaptionsoff
  \newpage
\fi

\section*{Acknowledgment}
This work is supported by Agile-IoT project (Grant No. 9131-00119B) granted by the Danish Council for Independent Research.

\bibliographystyle{IEEEtran}
\bibliography{ddNN}

\end{document}